\documentclass[%
 reprint,
 amsmath,amssymb,
 aps,
prl
]{revtex4-1}
\pdfoutput=1
\usepackage{graphicx}
\usepackage{dcolumn}
\usepackage{bm}
\usepackage{subfigure}

\DeclareRobustCommand{\orderof}{\ensuremath{\mathcal{O}}}

\begin{document}
\preprint{APS/123-QED}

\title{Axion Dark Matter Coupling to Resonant Photons via Magnetic Field}
\author{Ben T. McAllister}
\email{ben.mcallister@uwa.edu.au}
\author{Stephen R. Parker}
\affiliation{ARC Centre of Excellence for Engineered Quantum Systems, School of Physics, The University of Western Australia, 35 Stirling Highway, Crawley 6009, Western Australia, Australia}
\author{Michael E. Tobar}
\email{michael.tobar@uwa.edu.au}
\affiliation{ARC Centre of Excellence for Engineered Quantum Systems, School of Physics, The University of Western Australia, 35 Stirling Highway, Crawley 6009, Western Australia, Australia}
\date{\today}

\begin{abstract}
We show that the magnetic component of the photon field produced by dark matter axions via the two-photon coupling mechanism in a Sikivie Haloscope is an important parameter passed over in previous analysis and experiments. The interaction of the produced photons will be resonantly enhanced as long as they couple to the electric or magnetic mode structure of the Haloscope cavity. For typical Haloscope experiments the electric and magnetic coupling is the same and implicitly assumed in past sensitivity calculations. However, for future planned searches such as those at high frequency, which synchronize multiple cavities, the sensitivity will be altered due to different magnetic and electric couplings. We define the complete electromagnetic form factor and discuss its implications for current and future high and low mass axion searches, including some effects which have been overlooked, due to the assumption that the two couplings are the same.
\end{abstract}

\maketitle


Axions are a type of Weakly Interacting Slim Particle (WISP) originating from the Peccei Quinn solution to the strong CP problem in QCD~\cite{axion}. They can be formulated as highly motivated and compelling components of Cold Dark Matter (CDM)~\cite{Sikivie1983,Preskill1983,Dine1983,cdm}. Cosmological constraints provide upper and lower limits on the mass of the axion~\cite{limits}, yet still leave a large area of parameter space to be searched. One of the most mature and sensitive experiments is the Sikivie Haloscope~\cite{haloscope1,haloscope2}, which exploits the inverse Primakoff effect whereby a magnetic field provides a source of virtual photons in order to induce axion-to-photon conversion via a two photon coupling, with the generated real photon frequency being dictated by the axion mass. This signal is then resonantly enhanced by a cavity structure and resolved above the thermal noise of the measurement system. It has been well established that in a Haloscope with an axial DC magnetic field the expected power due to axion-to-photon conversion is given by~\cite{haloscope1,haloscope2,ADMX2011}
\begin{equation}
\text{P}_{a}=\left(\frac{g_{\gamma}\alpha}{\pi f_{a}}\right)^{2}\frac{\rho_{a}}{m_{a}}VB_{0}^{2}\text{Q}\text{C}, \label{eq:paxion}
\end{equation}
where $\text{g}_{\gamma}$ is a dimensionless model-dependent parameter of $\orderof(1)$~\cite{KSVZ1,KSVZ2,DFSZ1}, $\alpha$ the fine structure constant, $f_{a}$ the Peccei-Quinn energy breaking scale which dictates the axion mass and coupling strength, $m_{a}$ the axion mass, $\rho_{a}$ the local density of axions, $V$ the cavity volume, $B_{0}$ the applied magnetic field, Q the cavity quality factor (assuming the bandwidth is greater than the expected spread of the axion signal) and C the Haloscope form factor describing the overlap between the electric field created by the converted axions and the electric field structure of the resonant mode in the cavity.

To date Haloscope searches have excluded some areas of the parameter space~\cite{ADMX1,ADMX2011}, with further experiments currently under way and future efforts in various stages of planning~\cite{Baker2012,ADMXHF2014,wispdmx}. All of this work fundamentally relies on Eq.~\eqref{eq:paxion} to set constraints on $f_{a}$ and hence the mass of the axion and the strength of axion-photon coupling. In deriving Eq.~\eqref{eq:paxion} the coupling of an axion to a photon electric field is explicitly considered in the form factor, C, and it is then assumed that the corresponding magnetic field coupling is the same. Recent work describes the design of a magnetometer detection experiment enhanced by an LC circuit that utilizes the magnetic field coupling~\cite{LCaxions}. In this work we  consider the coupling of the photon magnetic field directly to a Haloscope cavity, and from this we are able to define the complete electromagnetic form factor for the axion Haloscope. Revising Eq.~\eqref{eq:paxion} to incorporate the complete form factor has major repercussions for the sensitivity of planned searches targeting high and low frequency axions, including those utilizing novel cavity geometries~\cite{Seviour2014,orpheus}.
\begin{figure}[t]
\centering
\includegraphics[width=0.95\columnwidth]{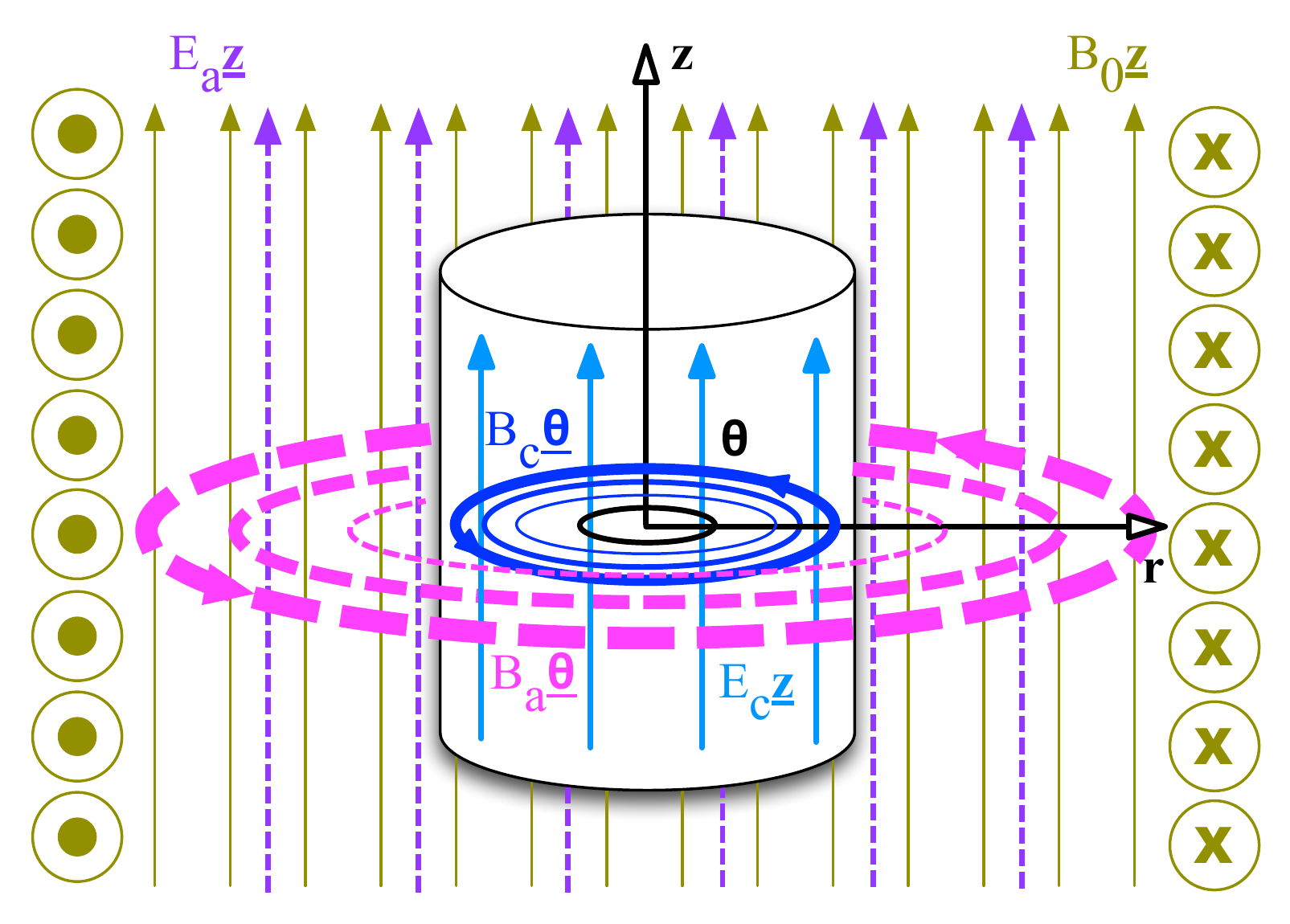}
\caption{Sketch of electromagnetic fields inside an axion Haloscope. The solenoid (gold) produces a static magnetic field, $\text{B}_0\vec{\hat z}$, which interacts with axions to produce an electric field (purple), $\text{E}_\text{a}\vec{\hat z}$ Eq.~\eqref{eq:Eax}, and a magnetic field (magneta), $\text{B}_\text{a}\vec{\hat \phi}$ Eq.~\eqref{eq:Bax}. For the TM$_{010}$ mode the cavity supports an electric field (aqua), $\text{E}_\text{c}\vec{\hat z}$, and a magnetic field (blue), $\text{B}_\text{c}\vec{\hat \phi}$.}
\label{fig:sketch}
\end{figure}

The modified Maxwell's equations accounting for an axion field, $a$, with no spatial dependence in the presence of a DC magnetic field as shown in fig.\ref{fig:sketch} are given by~\cite{haloscope1}
\begin{align}
\nabla\cdot\vec{E}&=0\qquad\nabla\cdot\vec{B}=0\nonumber\\
\nabla\times\vec{E}&=-\frac{\partial\vec{B}}{\partial t}\\
\nabla\times\vec{B}&=\frac{1}{c^{2}}\frac{\partial\vec{E}}{\partial t}-g_{\alpha\gamma\gamma}\frac{\vec{B_0}}{c}\frac{\partial a}{\partial t},\label{ME}
\end{align}
where $g_{\alpha\gamma\gamma}$ is the strength of axion-photon coupling (equal to $g_{\gamma}\alpha$/$\pi f_{a}$), $a=a_{0}e^{-j\omega_{a}t}$ and $\vec{B_0}=B_0\vec{\hat z}$ (the solenoid magnetic field). For this situation $\vec{E}=0$ and $\nabla\times\vec{B_0}=0$, thus Eq.~\eqref{ME} becomes:
\begin{align}
\nabla\times\vec{B_a}=\frac{1}{c^2}\frac{\partial\vec{E_a}}{\partial t},\label{eq:delBa}
\end{align}
where $\vec{B_a}$ is the magnetic field component of the photons produced via the axion-photon coupling. The right hand side of Eq.~\eqref{ME} defines the electric field generated by the axions as
\begin{align}
\vec{E_a}=-g_{\alpha\gamma\gamma}\,a\,c\,B_0\vec{\hat z}.\label{eq:Eax}
\end{align}

Applying Stokes Theorem inside the solenoid, from Eq.~\eqref{eq:delBa} and~\eqref{eq:Eax} we conclude that
\begin{align}
\vec{B_a}=-\frac{g_{\alpha\gamma\gamma}}{2c}rB_0\frac{\partial a}{\partial t}\vec{\hat\phi},\label{eq:Bax}
\end{align}
where r is the distance from the centre of the solenoid in the radial direction. 

\begin{figure*}[t]
\centering
\subfigure[]{
\includegraphics[width=0.45\textwidth]{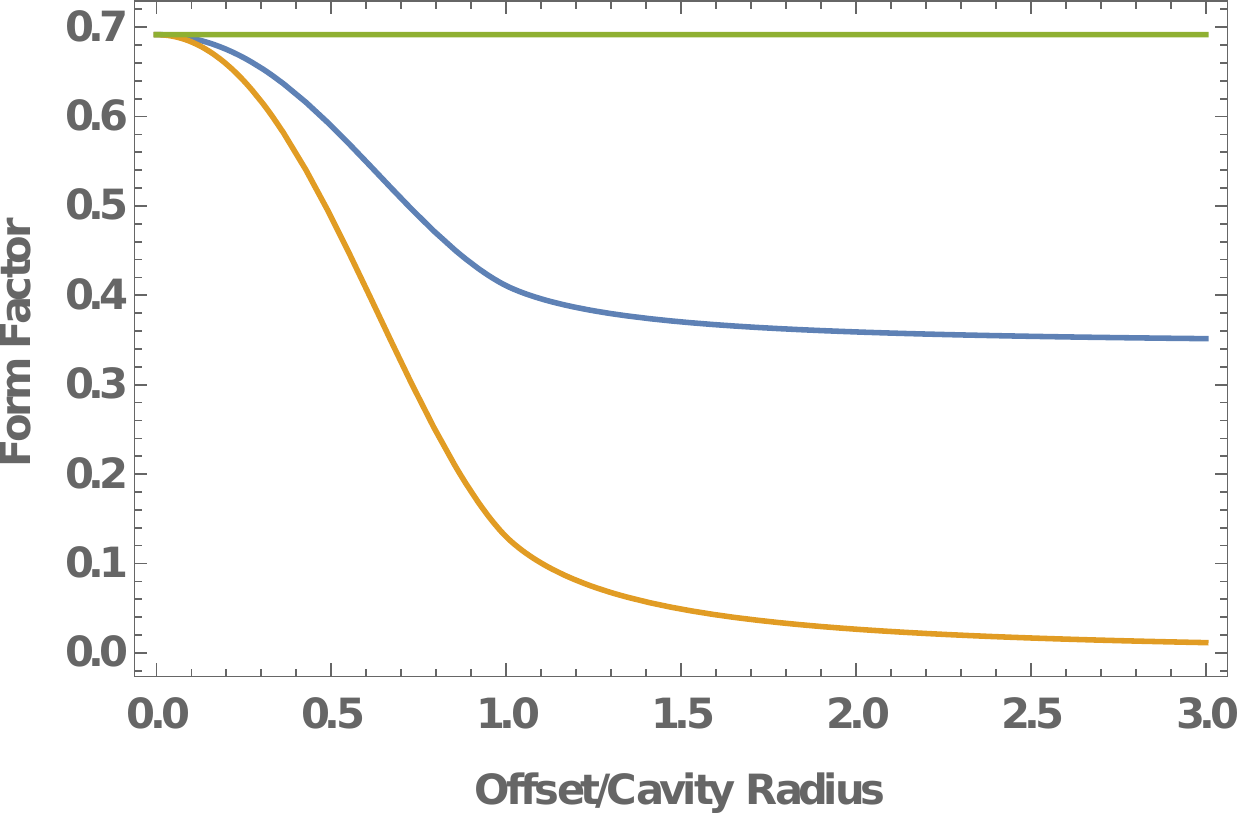}\label{fig:010}
}
\subfigure[]{
\includegraphics[width=0.45\textwidth]{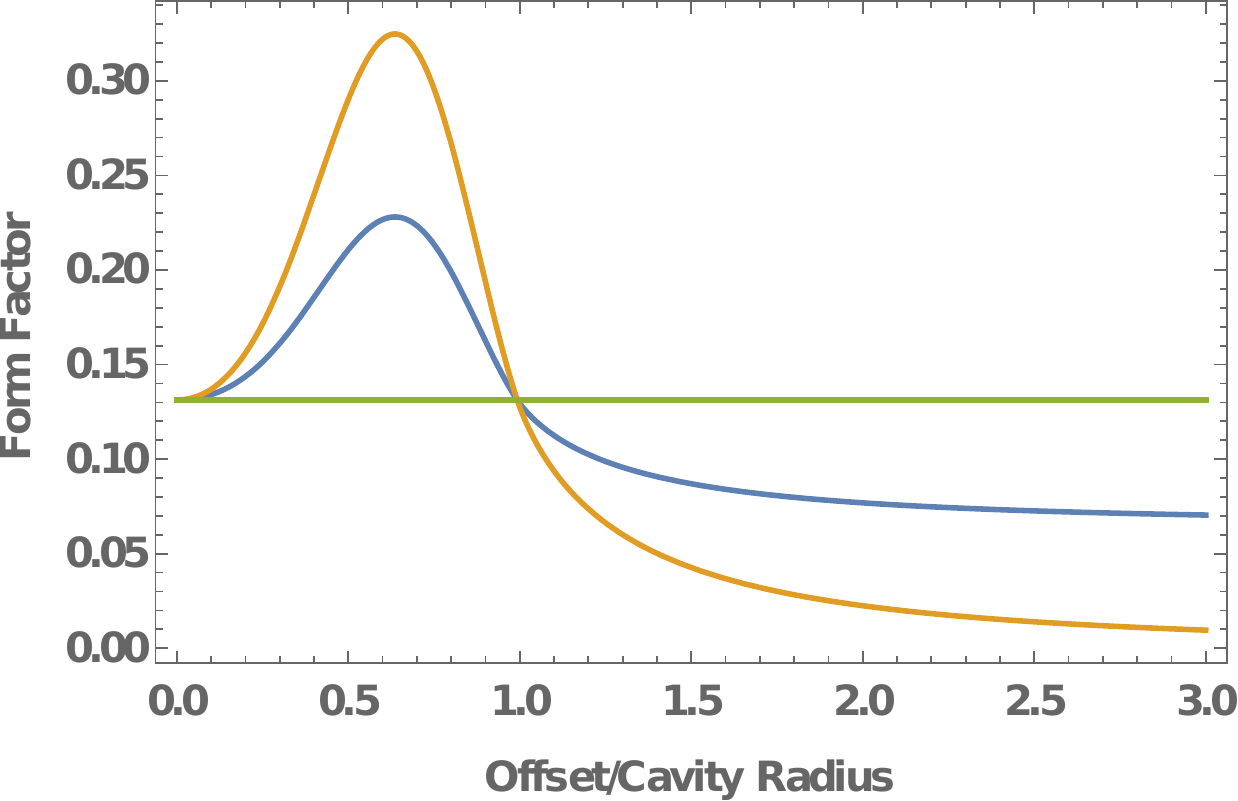}
}
\subfigure[]{
\includegraphics[width=0.45\textwidth]{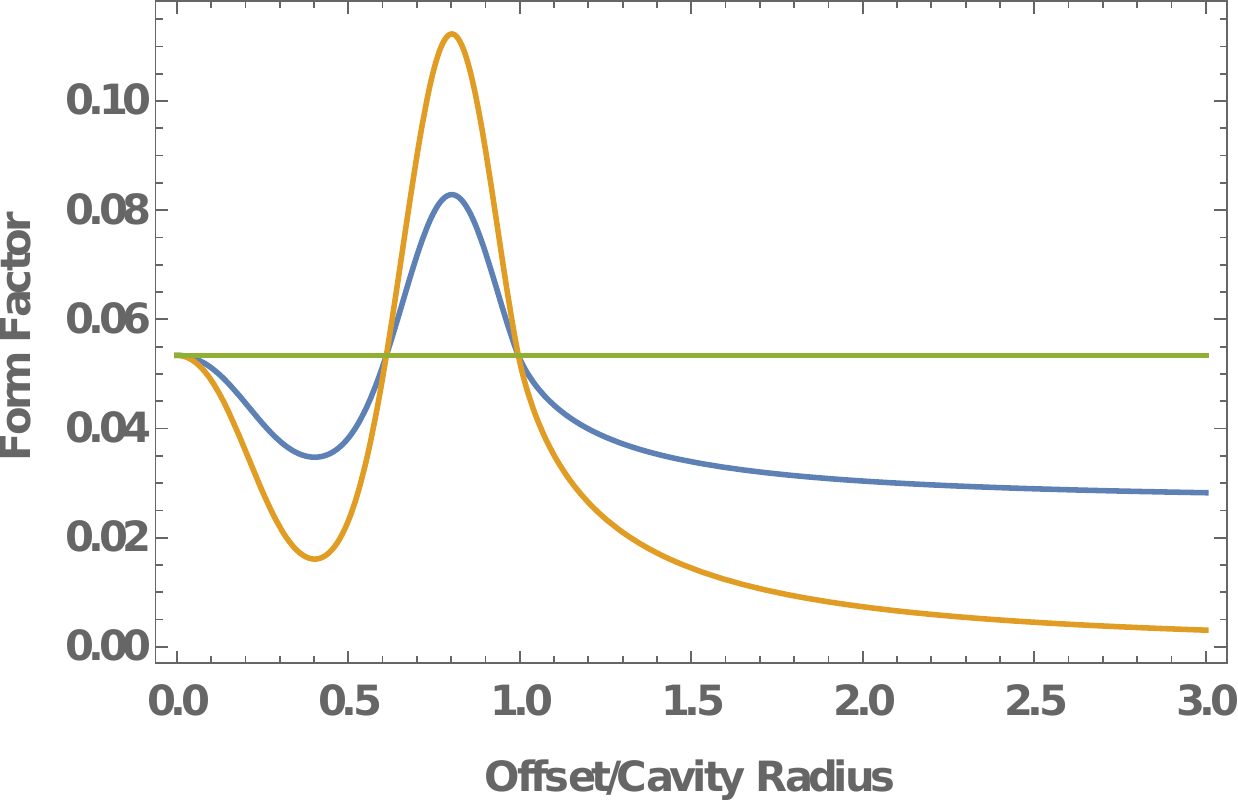}
}
\subfigure[]{
\includegraphics[width=0.45\textwidth]{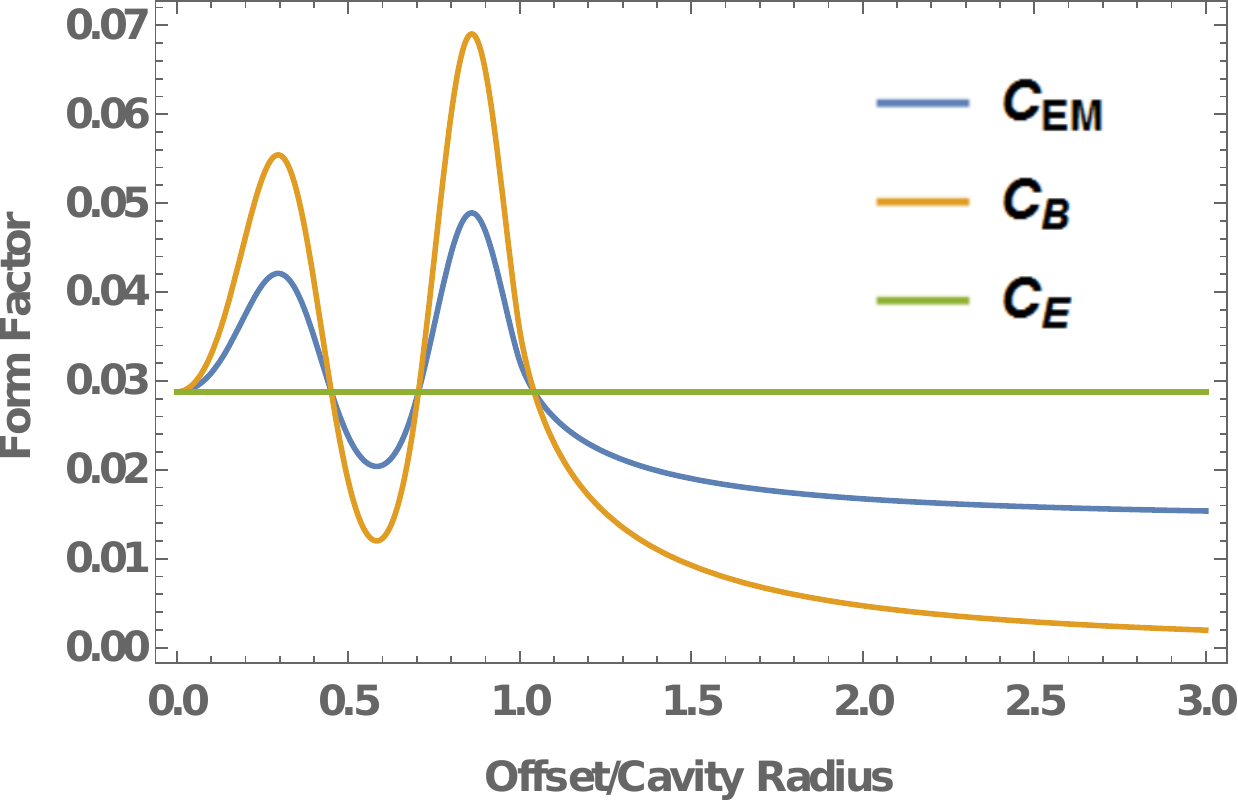}
}
\caption{Electric form factor (green), magnetic form factor (orange), and electromagnetic form factor (blue) as a function of the offset of the cavity centre from the solenoid centre, normalized to the cavity radius (e/a). Results are presented for the following modes: (a) TM$_{010}$, (b) TM$_{020}$, (c) TM$_{030}$ and (d) TM$_{040}$.}
\label{fig:FormOff}
\end{figure*}

We now consider a cylindrical cavity with resonant electric and magnetic mode field structures defined as $\vec{E_c}$ and $\vec{B_c}$ respectively. The electric energy stored in the cavity mode is given by
\begin{align}
\text{U}_e&=\frac{1}{4}\epsilon_0\int dV_{c}\mid E_c\mid ^2,\label{eq:EEn}
\end{align}
while the magnetic energy stored in the cavity mode is given by
\begin{align}
\text{U}_m&=\frac{1}{4}\frac{1}{\mu_0}\int dV_{c}\mid B_c\mid ^2.\label{eq:MEn}
\end{align}
Now we consider the electric and magnetic energy converted from axions. From the effective axion electric field (Eq.~\eqref{eq:Eax}) the electric energy converted into the cavity is
\begin{align}
\text{U}_{a,e}&=\frac{1}{4}\epsilon_0\,g_{\alpha\gamma\gamma}\,a\,c\,B_0\int dV_{c} \vec{E_c}\cdot\vec{\hat z},\label{eq:EaxEn}
\end{align}
while from the effective axion magnetic field (Eq.~\eqref{eq:Bax}) the magnetic energy converted into the cavity is
\begin{align}
\text{U}_{a,m}&=\frac{1}{4}\frac{1}{\mu_0}\frac{g_{\alpha\gamma\gamma}}{c}B_0\frac{\partial a}{\partial t}\int dV_{c}\,\frac{r}{2} \vec{B_c}\cdot\vec{\hat\phi}.\label{eq:BaxEn}
\end{align}
By equating the electric energy deposited into the cavity by the axions Eq.~\eqref{eq:EaxEn}, with the electric energy stored in the cavity mode, Eq.~\eqref{eq:EEn}, we can express the electric energy in the cavity due to axion conversion as
\begin{equation}
\text{U}_{a,e}=\frac{1}{2}c^{2}\epsilon_0\,g_{\alpha\gamma\gamma}^2\,a_\text{rms}^2\,B_0^2\,V\text{C}_\text{E},\label{eq:EEn2}
\end{equation}
where we have defined the cavity mode-dependent electric form factor as
\begin{align}
\text{C}_\text{E}=\frac{\left|\int dV_{c}\vec{E_c}\cdot\vec{\hat z}\right|^2}{V\int dV_{c}\mid E_c\mid^2}.\label{CE}
\end{align}
This is the standard axion Haloscope form factor commonly denoted by $C$ throughout the literature (see Eq.~\eqref{eq:paxion}). It has been implicitly assumed that the magnetic field contributes the same amount of energy and so a factor of two has been applied to Eq.~\eqref{eq:EEn2} in the past to calculate the sensitivity of Haloscope experiments. Now, in this work we explicitly consider the magnetic field contribution. Proceeding as before, by considering Eq.~\eqref{eq:MEn} and~\eqref{eq:BaxEn} we can express the magnetic energy in the cavity due to axion conversion as
\begin{align}
\text{U}_{a,m}&=\frac{1}{2}\frac{1}{\mu_0}\,g_{\alpha\gamma\gamma}^2\,a_\text{rms}^2\,B_0^2\,V\text{C}_\text{B},\label{eq:BEn2}
\end{align}
where we define the cavity mode-dependent magnetic form factor as
\begin{align}
\text{C}_\text{B}=\frac{\frac{\omega_a^2}{c^2}\left|\int dV_{c}\frac{r}{2}\vec{B_c}\cdot\vec{\hat\phi}\right|^2}{V\int dV_{c}\mid B_c\mid^2}.\label{CB}
\end{align}
It is worth emphasizing that $r$ and $\phi$ refer to the solenoid coordinates, not the cavity. The total electromagnetic energy stored in the cavity will be the sum of the electric and magnetic contributions given by Eq.~\eqref{eq:EEn2} and Eq.~\eqref{eq:BEn2}
\begin{align}
\text{U}_{a,em}=\text{U}_{a,e}+\text{U}_{a,m}=\frac{1}{2}\frac{1}{\mu_0}g_{\alpha\gamma\gamma}^2\,B_0^2\,V\,a_{rms}^2(\text{C}_\text{E}+\text{C}_\text{B}).\label{eq:EBEn}
\end{align}
To obtain the signal power expected on resonance, we multiply Eq.~\eqref{eq:EBEn} by $\omega_a \text{Q}_\text{L}$ such that
\begin{align}
\text{P}_{a}&=\frac{1}{\mu_0}g_{\alpha\gamma\gamma}^2\,a_{rms}^2\omega_a\,V\,B_0^2\text{Q}_\text{L}\text{C}_\text{EM} \nonumber\\
&=g_{\alpha\gamma\gamma}^2\frac{\rho_{a}}{m_{a}}\,V\,B_0^2\text{Q}_\text{L}\text{C}_\text{EM},
\label{halo}
\end{align}
(see \cite{dawthesis}) here the axion frequency, $\omega_{a}$, is equal to the cavity resonance frequency and C$_{\text{EM}}$ is the complete electromagnetic form factor, $\left(\text{C}_{\text{E}}+\text{C}_{\text{B}}\right)/2$. For the standard Haloscope design, where the cavity resonant TM mode is axially symmetric with respect to the central axis of the applied magnetic field, the electric and magnetic form factors are equal (C$_{\text{E}}$=C$_{\text{B}}$) and  Eq.~\eqref{halo} reverts to Eq.~\eqref{eq:paxion}. However, due to the $r$ dependence in C$_{\text{B}}$ (Eq.~\eqref{CB}) it is clear that for more general experiments with cavities offset from axial symmetry, or with separated electric and magnetic fields, the value of $\text{C}_{\text{E}}$ and $\text{C}_{\text{B}}$ would not be equal, and the complete electromagnetic form factor would need to be calculated. For example, efforts to search for higher frequency axions can involve power-summing multiple cavities within a solenoid~\cite{multicav,ADMXHF2014}; each of these nominally identical cavities will produce different amounts of power depending upon their relative location within the solenoid. Now we explore the sensitivity of a single resonant cavity offset from the centre of a solenoid, which obviously implies that the radius of the cavity is smaller than that of the solenoid. Such a setup is most applicable to higher frequency axion searches, either with a single cavity or multiple cavities. Considering the radial dependence of the magnetic form factor defined in Eq.~\eqref{CB}, and understanding that this relates to the radial value of the solenoid's natural cylindrical coordinate system, we must perform a coordinate transform in order for the dot product of the axion and cavity fields to be physically meaningful. This gives the following expression for the magnetic form factor,
\begin{align}
\text{C}_{\text{B}}=\frac{\frac{\omega_a^2}{c^2}\left|\int dV_{c}\text{B}_{\text{c}_\phi}\frac{\text{r}-\text{e}\cos{\phi}}{2}\right|^2}{V\int dV_{c}\mid B_c\mid^2},\label{eq:CBcomp}
\end{align}
where $\text{B}_{\text{c}_\phi}$ is the $\phi$ component of the cavity magnetic field, r and $\phi$ are the solenoid's natural radial and azimuthal coordinates and e is the offset of the centre of the cavity from the centre of the solenoid. We could perform a similar transformation for the electric field, but for the TM$_{0x0}$ mode family as employed by Haloscope experiments it is trivial as the z directions of the two systems are the same. Form factors for TM$_{0x0}$ modes were computed numerically as a function of e/a, where a is the cavity radius. Fig. \ref{fig:FormOff} shows plots of the magnetic form factor, $\text{C}_{\text{B}}$, the electric form factor, $\text{C}_{\text{E}}$, and the combined electromagnetic form factor, C$_{\text{EM}}$ ($=\frac{\text{C}_{\text{B}}+\text{C}_{\text{E}}}{2}$), for various TM modes as a function of e/a. While the traditional electric form factor remains constant for all modes as expected, the electromagnetic form factor decreases with offset from the centre for the $\text{TM}_{010}$ mode, but for higher order modes we see positions with an increased form factor. For a $\text{TM}_{010}$ mode, as the cavity moves away from the centre of the solenoid, some of the $\phi$ direction magnetic field of the cavity, which was previously in phase with the solenoid field is now opposing the magnetic field of the solenoid (see fig.~\ref{fig:sketch}). The cancellation of fields leads to reduced mode overlap and lower sensitivity. However, for higher order modes with alternating in and out of phase magnetic field components, as the cavity is offset from the centre there are regions where more of the cavity magnetic field is in the same direction as the solenoid field (as it was previously in the opposite direction when the cavity was central), thus achieving a higher overlap and a higher form factor. The greatest improvement can be seen in the $\text{TM}_{020}$ mode, with a sensitivity increase of ~75$\%$. Clearly, great care must be taken when positioning cavities within the solenoid to avoid reduced sensitivity to axions. Conversely, this also opens up the potential to increase the sensitivity of Haloscope axion searches, which use higher order modes. Whilst the axion magnetic coupling has been considered before, the implication of this coupling for many current and future proposed axion searches has not been explicitly treated. Any cavity-based axion detection scheme must consider this coupling, as even a slight deviation from perfect central placement of cavities in a magnetic field will effect the complete electromagnetic form factor. Furthermore, it is often assumed that metal tuning rods which are commonly employed in such searches do not significantly alter the mode overlap - typically only the electric coupling is considered in such discussions and it is clear that the magnetic coupling will be altered in a different way. Any experiments which introduce dielectric media into the cavity volume will also need to consider the impact of these dielectrics on the electric and magnetic form factors explicitly. Finally, detection of axions via the magnetic coupling has been proposed through the use of LC circuits as the readout mechanism~\cite{LCaxions} - as we have now defined the complete electromagnetic form factor for cavity-based experiments, we have opened the possibility of low mass, preinflationary~\cite{preinflation} axion detection via 3D lumped LC resonators, commonly known as re-entrant cavities~\cite{RRcav,rigorous,tuning,splitring}. In conclusion, an axion in the presence of a magnetic field will convert into a real photon with electric and magnetic field components. We have shown how the generated magnetic field component interacts with a resonant cavity structure, such as those utilized in axion Haloscope searches. In doing so we have defined the complete electromagnetic form factor for axion Haloscopes, which has major repercussions for the sensitivity of future axion experiments searching for both low and high mass axions.\\
\\
The authors thank Ian McArthur for useful discussions. This work was supported by Australian Research Council grants CE110001013 and DP130100205.

\end{document}